\documentclass{article}
\usepackage[utf8]{inputenc}
\usepackage{amsmath}
\usepackage{graphicx}
\usepackage{authblk}
\usepackage{hyperref}
\usepackage{subcaption}
\usepackage{float}
\usepackage{adjustbox}

\title{\LARGE EchoVest: Real-Time Sound Classification and Depth Perception Expressed through Transcutaneous Electrical Nerve Stimulation}
\author{Jesse Choe, Siddhant Sood, and Ryan Park}
\date{}

\begin{document}

\maketitle

\begin{abstract}
Over 1.5 billion people worldwide live with hearing impairment [17]. Despite various technologies that have been created for individuals with such disabilities, most of these technologies are either extremely expensive or inaccessible for everyday use in low-medium income countries. In order to combat this issue, we have developed a new assistive device, EchoVest, for blind/deaf people to intuitively become more aware of their environment. EchoVest transmits vibrations to the user’s body by utilizing transcutaneous electric nerve stimulation (TENS) based on the source of the sounds. EchoVest also provides various features, including sound localization, sound classification, noise reduction, and depth perception. We aimed to outperform CNN-based machine-learning models, the most commonly used machine learning model for classification tasks, in accuracy and computational costs. To do so, we developed and employed a novel audio pipeline that adapts the Audio Spectrogram Transformer (AST) model, an attention-based model, for our sound classification purposes, and Fast Fourier Transforms for noise reduction. The application of Otsu’s Method helped us find the optimal thresholds for background noise sound filtering and gave us much greater accuracy. In order to calculate direction and depth accurately, we applied Complex Time Difference of Arrival algorithms and SOTA localization. Our last improvement was to use blind source separation to make our algorithms applicable to multiple microphone inputs. The final algorithm achieved state-of-the-art results on numerous checkpoints, including a 95.7\% accuracy on the ESC-50 dataset for environmental sound classification.
\end{abstract}

\section{Introduction}
According to the World Health Organization, if a person's hearing thresholds are below 20 dB, they are said to have hearing loss [17]. The consequences of this condition can vary in their severity and may include difficulties in communication, leading to social isolation for older individuals, decreased academic performance in children, and limited job opportunities for adults in areas without adequate accommodations for those with hearing loss. Currently, 1.5 billion people globally, or one in every five people, live with hearing loss and this is projected to increase to 2.5 billion people, or 25\%, of the world population by 2050. The majority of individuals with hearing loss, 80\%, live in low and middle-income countries. Despite this, a significant amount of hearing loss goes unaddressed, costing governments around the world nearly \$980 billion annually, with the majority of these costs incurred in low and middle-income countries. This high cost is attributed to the expensive nature of hearing impairment devices such as hearing aids and cochlear implants, which range in cost from \$2,000 to \$7,000 for hearing aids [16] and \$30,000 to \$50,000 for cochlear implants [10].

We aim to address the problem of unaddressed hearing impairment by creating EchoVest, a cost-effective wearable alternative to current hearing impairment solutions. EchoVest utilizes sound localization and depth perception through the use of TENS pads and sound classification through Audio Spectrogram Transformers (ASTs). To implement sound localization and depth perception, EchoVest selectively activates TENS pads [14] with amplified signals based on the distance and location of the sound source. With a total cost of manufacturing of \$98.90, EchoVest is a significantly more affordable and effective option than existing hearing impairment technologies.

\section{Materials}
The primary objective was to create an inexpensive, durable, and wearable device for the user’s day to day activities. We used a mesh vest as the base with wiring interwoven throughout the mesh. A mesh vest is ideal for contact between skin and the output nodes. On the back of the mesh, we employed a Raspberry Pi 3B+ as our central computer in order to control all of the output devices. The Machine-Learning libraries and other built-in software that we integrated into our algorithms all required a 64-bit system and the Raspberry Pi 3B+ was the cheapest processor on the market that we were able to obtain. In order to record sounds, we used a ReSpeaker 4-mic Array due to the fact that we could get 4 different streams of audio input. The continuous stream of 4 mics made it possible for us to calculate the direction and distance of sounds. The last piece of significant hardware that we used were TENS electrodes. TENS is a service that delivers mild electrical currents through electrodes placed on the body. By applying TENS to our vest, we are able to directly stimulate the user’s nerves and leave them with a multi-dimensional feeling. A full materials list breakdown with all other assorted materials can be seen in the Figure 1 below. Our essential pieces of hardware and their application can be seen in Figure 2.

\begin{figure}[H]
  \centering
  \begin{subfigure}[b]{0.45\textwidth}
    \includegraphics[width=\textwidth]{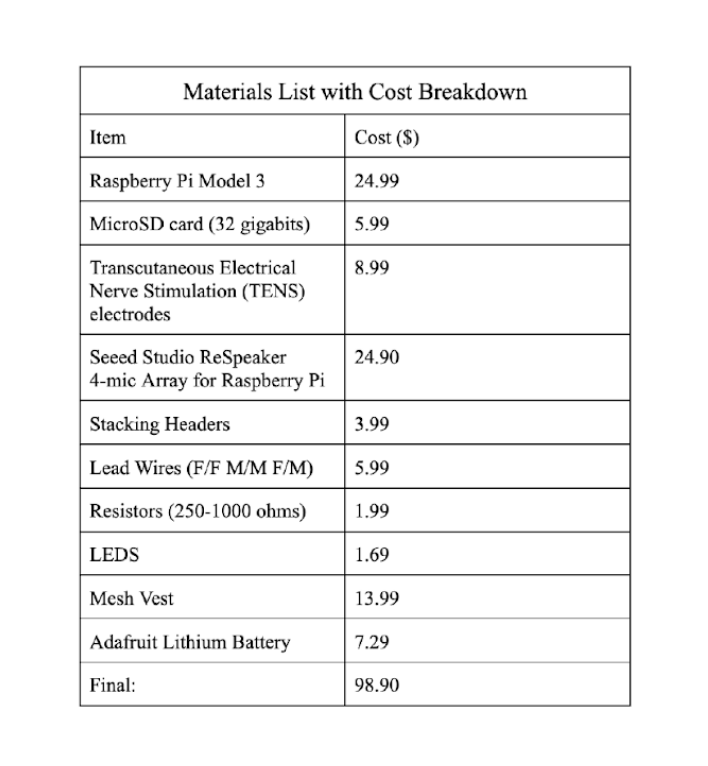}
    \captionsetup{labelformat=empty} % Removes label
    \caption{Figure 1: Materials List with Costs}
    \label{fig:image1}
  \end{subfigure}
  \hfill
  \begin{subfigure}[b]{0.5\textwidth}
    \includegraphics[width=\textwidth]{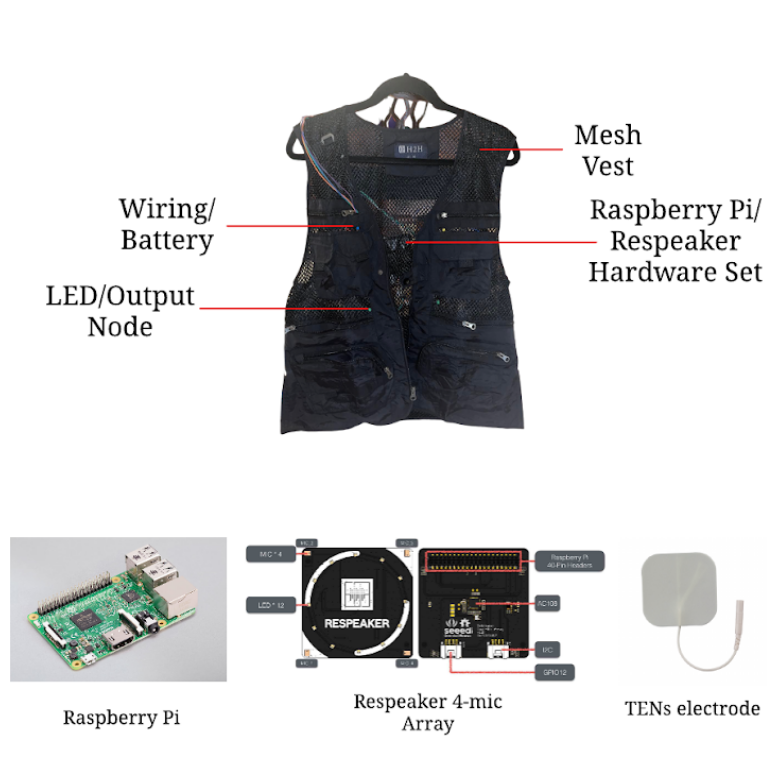}
    \captionsetup{labelformat=empty} % Removes label
    \caption{Figure 2: Materials Diagram on Vest}
    \label{fig:image2}
  \end{subfigure}
  \label{fig:main}
\end{figure}

\section{Methods}

We designed EchoVest to determine the relative location and distance of a sound source from each microphone in our vest for audio spatial awareness. We were able to triangulate the relative location of the sound source in real time and determine the sound's arrival angle using the Open embeddeD Auditory System's (ODAS) built-in sound localization algorithms [4]. We utilized Time Difference of Arrival (TDoA) with Generalized Cross-Correlation with Phase Transforms (GCC-PHAT) to calculate the distance from the sound source to the microphones in real time. After recording the audio signals coming from each microphone, we calculated the cross-correlation function by sliding one signal in relation to the other for each time step to see how similar their waveforms are to one another. The time delay between the signals, or TDoA value, between the two microphones is represented by the time step with the highest cross-correlation value [8]. We were able to select the appropriate pad based on the angle of sound arrival and alter the strength of our electrical pad signals in response to the distance from the microphone, calculated using the distance-rate-time equation.

In order to enhance EchoVest's sound localization and depth perception, we utilized a Blind Source Separation (BSS) approach that combined Principal Component Analysis (PCA), Non-Negative Matrix Factorization (NMF), and Independent Component Analysis (ICA) to separate the combined sound file from the microphone array into individual sound files for each microphone. We first reduced the dimensionality of the sound input with PCA and NMF. NMF factorized the sound input into two smaller matrices, as seen in Figure 3, with non-negative elements [15], while PCA transformed the sound data into a new coordinate system with axes along the directions of maximum variance [7]. ICA then separated the mixed sound signal into subcomponents by assuming that only one component was Gaussian and that the components were independent from each other [4]. The cross-correlation matrix and TDOA values from sound localization allowed us to identify each sound with its corresponding microphone, resulting in enhanced sound localization and depth perception.

\begin{figure}[H]
  \centering
  \includegraphics[width=\textwidth]{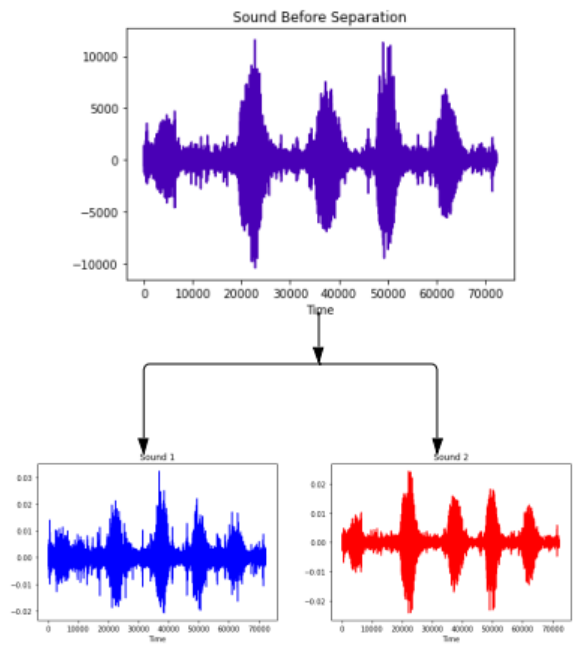}
  \captionsetup{labelformat=empty}
  \caption{Figure 3: The original sound (top) was separated into two separate sounds (left and right)}
  \label{fig:image3}
\end{figure}

We employed a signal processing approach that combined Fast Fourier Transforms (FFTs) with Otsu's method to effectively remove background noise from our sound input and enhance the sound classification accuracy. First, we converted the sound input into its frequency domain using FFT, an efficient algorithm that computes the discrete Fourier transform in real-time. We then implemented Otsu's Method, a commonly used image processing algorithm, to denoise the sound input by selecting an optimal noise threshold. Otsu's Method classifies pixels into background and foreground based on their intensity levels and was applied to the audio by converting it into a frequency histogram from the Fourier transform [2]. This effectively filtered out white noise from the input signals and improved the sound classification accuracy. Figures comparing the sound before and after applying Fast Fourier Transform (FFT) with Otsu's Method are presented below in Figure 4.

\begin{figure}[H]
  \centering
  \includegraphics[width=\textwidth]{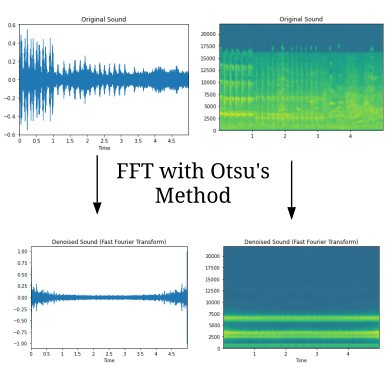}
  \captionsetup{labelformat=empty}
  \caption{Figure 4: The sound waves before and after noise reduction (left) and the audio spectrograms before and after noise reduction (right)}
  \label{fig:image4}
\end{figure}

We implemented a Sound Spectrogram Transformer that was trained on the ESC-50 dataset, a dataset for Environmental Sound Classification which consists of 2000 environmental audio recordings suitable for benchmarking methods of environmental sound classification, using 5-fold cross-validation to prevent overfitting. Additionally, this dataset was used in order to limit the amount of semantical classes to only classify sounds frequently associated with real-time environments. As shown in Figure 5, this transformer takes the audio waveform input of T seconds, outputted from Otsu’s Method, and is converted into a 128x100t spectrogram using log Mel filterbank features computed with a 25ms Hamming window every 10ms. The model outputs a Transformer encoder's [CLS] token, which serves as the audio spectrogram representation for classification. The corresponding label is matched by using a linear layer with sigmoid activation.   

To assess the accuracy of the classification data produced by the transformer model in a real-time environment, the resultant semantic class was paired with a timestamp associated with the live sound data. This live sound data was simulated by playing a variety of sounds around the mic array with white noise that included people talking, laughter, and the air conditioner running. The aligned time series data were then used to calculate errors and determine accuracy of the real-time classification system. The sample capture rate for the audio input was 0.205 seconds. 

\begin{figure}[H]
  \centering
  \includegraphics[width=\textwidth]{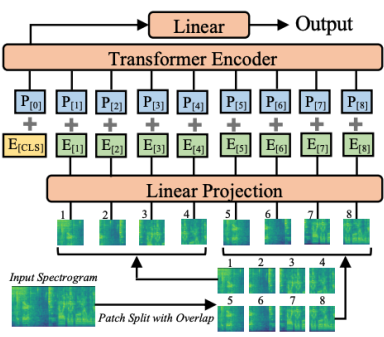}
  \captionsetup{labelformat=empty}
  \caption{Figure 5: Audio Spectrogram Transformer (AST) Model Architecture }
  \label{fig:image5}
\end{figure}

The only constraint for the electrical system was that we needed to provide an equal amount of current and voltage to each of the output nodes located on the vest. The intensity of the electrical current expelled from the TENS electrodes is directly related to the current from the Raspberry Pi, which outputs a maximum current of 50 hertz through the 5V pin-outs. By applying two strategic parallel circuits, as demonstrated in Figure 6, we were able to ensure that each output node only revised a maximum current of 12.5 HZ. 12.5 HZ is under-powered for the typical TENS electrode (50 Hertz), the current is still high enough to be felt and also guarantees user safety.

\begin{figure}[H]
  \centering
  \includegraphics[width=\textwidth]{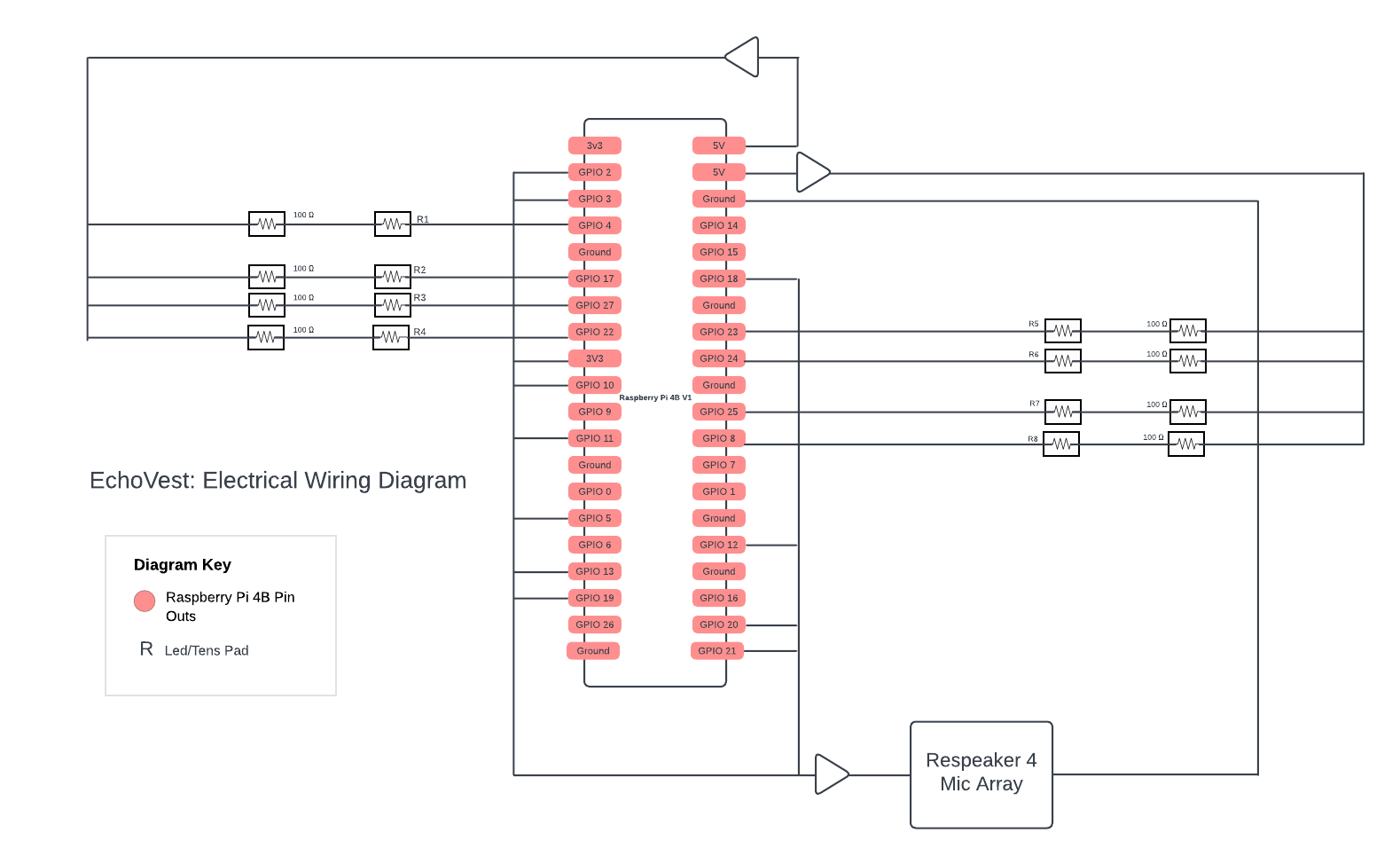}
  \captionsetup{labelformat=empty}
  \caption{Figure 6: Electrical Diagram}
  \label{fig:image6}
\end{figure}

The process for setting up the OS and driver was challenging due to many libraries being outdated and the Re-Speaker 4 Mic Array Drivers being meant for a 32-bit version of Raspberry Pi. We downloaded specific versions of packages and downgraded our Raspberry Pi OS to 64-bit Raspberry Pi OS Bullseye 11.0 Debian Release. Lastly, the Re-Speaker 4-Mic Array Drivers was written for a 32-bit system with Linux Kernel Version 4.9.80+ while the earliest release of Raspberry Pi 64-bit OS had 5.10.40+ Kernels. We addressed this issue by modifying the ReSpeaker Driver Scripts and creating overlay diversions for every component of the driver.

\section{Results}

\begin{figure}[htbp]
  \centering
  \includegraphics[width=0.5\textwidth]{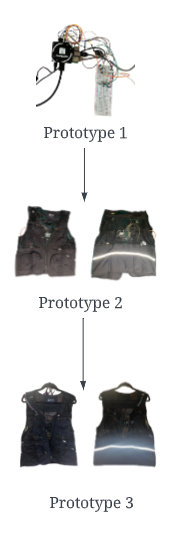}
  \captionsetup{labelformat=empty}
  \caption{Figure 7 : Prototype Development}
  \label{fig:image7}
\end{figure}

Referencing Figure 7, Prototype 1 had a simple design which served as a basic setup on a breadboard, comprising the ReSpeaker, LEDs, and the Raspberry Pi. This prototype allowed us to test our ML algorithms and hardware, which enabled accurate node activations and the assessment of our model's accuracy in the presence of live background noise. The next stage of prototyping involved integrating our breadboard design with the mesh vest, to test its practical functionality and identify any issues with our original mesh design. Prototype 3 was focused on making the vest portable by incorporating a battery pack, streamlining the wiring, and correcting node positioning. Additionally, the position of the ReSpeaker was adjusted to eliminate significant vest feedback that was observed in the previous prototype. As a result, the device is now easy to use, simply by putting on the vest and switching on the battery pack. We tested the device with LEDS by having a test subject put on the vest and had a person playing sounds on a phone from 2 meters away. By visually being able to see the changes in luminosity of the LEDS, we could confirm that our depth and direction algorithms were working.

As shown in Table 1, the implementation of Fast Fourier Transforms with Otsu's Method outperformed three other noise reduction techniques with a peak signal-to-noise ratio (PSNR) of 57.5 dB. FFT with Otsu's Method is an uncommon technique for noise reduction, but it proved to be more effective than each of the other algorithms due to its higher PSNR value, which indicates the algorithm's ability to reduce noise. Referencing Table 2, our Audio Spectrogram Transformer model outperformed many traditional sound classification models in terms of accuracy, with higher accuracies on the ESC-50 dataset and higher mean average precisions (mAPs) on the AudioSet dataset. Specifically, the Audio Spectrogram Transformer achieved an accuracy of 95.7\% on the ESC-50 dataset and a 0.485 mAP on the AudioSet.

\begin{table}[H]
  \centering
  \caption{Comparison of PSNR Values for Various Noise Reduction Algorithms}
  \label{tab:psnr}
  \begin{tabular}{|c|c|}
    \hline
    \textbf{Algorithm} & \textbf{PSNR (dB)} \\
    \hline
    Wiener Filtering & 36.791 \\
    Spectral Gating & 55.235 \\
    Spectral Subtraction & 57.116 \\
    \textbf{FFT with Otsu's Method} & \textbf{57.529} \\
    \hline
  \end{tabular}
  
  \medskip
  \footnotesize
  Note: Peak signal-to-noise ratio (PSNR) measures the amount of noise present in a signal and how it affects the quality of the signal. It is expressed in decibels (dB), and a higher PSNR value indicates a lower amount of noise and a higher quality signal.
\end{table}

\begin{table}[H]
  \centering
  \caption{Table 2: Accuracy Comparison of Various Sound Classification Models on the ESC-50 Dataset}
  \label{tab:psnr}
  \begin{tabular}{|c|c|}
    \hline
    \textbf{Model} & \textbf{Accuracy (\%)} \\
    \hline
    auDeep & 64.3 \\
    WSNet & 66.25 \\
    SoundNet & 74.2 \\
    Human & 81.3 \\
    AclNet & 85.65 \\
    AVID & 89.2 \\
    EAT-S & 95.25 \\
    \textbf{AST} & \textbf{95.7} \\
    \hline
  \end{tabular}
\end{table}

Also, we conducted a small test of the electrical stimulation of the TENS electrodes on the vest. We had 10 different human volunteers (55-65 years old) report the amount of stimulation that they received on a scale from 0-10 (with 0 being no electrical stimulation and 10 being the stimulation with the maximum current output) when all the TENS were activated. This process was then repeated in 3 different environments.

\begin{table}[H]
  \centering
  \renewcommand{\arraystretch}{1.2}
  \caption{The Effect of Different Current levels from a TENS Electrode and Changes of Environment on the Physical Stimulation Received (0-10)}
  \label{tab:results}
  \begin{adjustbox}{center}
  \begin{tabular}{|c|*{9}{c|}}
    \hline
    \multicolumn{1}{|c|}{\textbf{Testing Volunteer \#}} & \multicolumn{9}{c|}{\textbf{Environment}} \\
    \cline{2-10}
    \multicolumn{1}{|c|}{} & \multicolumn{3}{c|}{Indoors at 68°F (control)} & \multicolumn{3}{c|}{Outdoors at 68°F} & \multicolumn{3}{c|}{Outdoors at 32°F} \\
    \cline{2-10}
    \multicolumn{1}{|c|}{\textbf{Current (mA)}} & 0 & 6 & 12.5 & 0 & 6 & 12.5 & 0 & 6 & 12.5 \\
    \hline
    1 & 0 & 5 & 10 & 0 & 6 & 10 & 0 & 5 & 10 \\
    2 & 0 & 6 & 10 & 1 & 6 & 10 & 1 & 6 & 10 \\
    3 & 0 & 5 & 10 & 0 & 5 & 10 & 1 & 5 & 10 \\
    4 & 0 & 5 & 10 & 0 & 5 & 10 & 0 & 5 & 10 \\
    5 & 0 & 6 & 10 & 0 & 5 & 10 & 1 & 5 & 10 \\
    6 & 0 & 5 & 10 & 0 & 5 & 9 & 0 & 6 & 10 \\
    7 & 0 & 6 & 10 & 0 & 5 & 10 & 0 & 5 & 10 \\
    8 & 0 & 6 & 10 & 0 & 5 & 10 & 0 & 6 & 9 \\
    9 & 0 & 6 & 10 & 0 & 5 & 10 & 0 & 6 & 10 \\
    10 & 0 & 6 & 10 & 0 & 5 & 10 & 0 & 6 & 10 \\
    \hline
  \end{tabular}
  \end{adjustbox}
\end{table}

\section{Discussion}
Our implementation of the Fast Fourier Transform (FFT) algorithm using Otsu's Method for noise thresholding effectively removed white noise from our audio input, which preserved the model’s sound classification accuracy of 95.7\%. This confirmed that our preprocessing pipeline was efficient in accurate sound classification in real-time environments. This implementation was much more effective than other methods, since the peak signal-to-noise ratio (PSNR) was higher than other methods without requiring machine learning, which allowed our algorithm to process under the limited computational powers. We tested the different noise reduction algorithms in Table 1 by calculating the PSNR (using the PSNR formula) of the original and denoised sounds for each of the four different noise reduction algorithms we considered. 

As shown in Table 3, the TENS stimulation is largely unaffected by changes in temperature and is largely unaffected by environmental factors. Due to a lack of professional expertise and knowledge of wiring and Raspberry Pi current control, we could not guarantee the safety of the user due to the possible rapid changes in current. Thus, we replaced the TENS with LEDs, which were the closest alternative to the TENS as they could describe the distance and depth through visual aid with the brightness of the LED indicating distance and the specific placement of the LED representing direction. As a result, EchoVest serves as a proof-of-concept that the various components and pipeline accomplish the task. 

Currently, we are creating an app that hosts the classification model and sound preprocessing on the cloud because of potential for customizable features. However, EchoVest will still be completely localized and will not require the app to function. The app serves as a means to increase functionality with wifi and bluetooth capabilities by giving the user the ability to not be notified of certain sounds and change the various strengths of the TENS to the person's suitability. Furthermore, our future goal includes a partnership with companies like Ring that would allow us to utilize and implement our sound classification and preprocessing pipeline to their system. Their systems are outdated and do not take out sufficient white noise and, therefore, limit the functionality of their doorbell system. In addition to company partnerships, EchoVest can be directly applied to search and rescue operations as it gives personnel a heightened sense of their surroundings. EchoVest has multiple functionalities, which makes EchoVest a multi-purpose solution to a variety of real world problems.

\section{Conclusion}
In this paper, we developed an accessible, cost-efficient wearable product for localizing and classifying sounds. We further demonstrate that our preprocessing pipeline, consisting of Otsu’s Method and FFT, sufficiently preserves sound classification accuracy in a real-time environment. EchoVest utilizes optimized machine learning to efficiently and effectively lower computation costs, thereby reducing product costs. EchoVest costs a maximum of \$98 per unit to manufacture and is easily accessible to the general public due to the lack of customization needed. Traditional hearing aid devices require numerous prerequisites, such as hearing test, medical clearance, and hearing aid evaluation. Constrastingly, EchoVest will be available to the public without any specialization or pre-requisites because its only variable would be the size of the vest.

\end{document}